\documentclass[aps,showpacs,preprint,preprintnumber,nofootinbib,amsmath,amssymb,12pt]{revtex4}
\usepackage{bm}
\usepackage[dvipdfmx]{graphicx,color}

\begin{document}

\title{\vspace*{0.5cm}
Charged black strings in a five-dimensional Kasner universe
}
\bigskip
\author{
${}^{1}$Hideki Ishihara\footnote{E-mail: ishihara@sci.osaka-cu.ac.jp}, 
${}^{2}$Masashi Kimura\footnote{E-mail: m.kimura@damtp.cam.ac.uk}
and
${}^{1}$Ken Matsuno\footnote{E-mail: matsuno@sci.osaka-cu.ac.jp}
\bigskip
\bigskip
}
\affiliation{
${}^{1}$Department of Mathematics and Physics, Osaka City University, Sumiyoshi, Osaka 558-8585, Japan
\\
${}^{2}$DAMTP, University of Cambridge, Centre for Mathematical Sciences,
Wilberforce Road, Cambridge CB3 0WA, UK
\bigskip
\bigskip
}

\begin{abstract}
We construct time-dependent charged black string solutions 
in five-dimensional Einstein-Maxwell theory. In the far region, 
the spacetime approaches a five-dimensional Kasner universe
with an expanding three-dimensional space and a shrinking extra dimension. 
Near the event horizon, 
the spacetime is approximately static and has a smooth event horizon.
We also study the motion of test particles around the black string
and show the existence of quasi-circular orbits. Finally, we briefly 
discuss the stability of this spacetime.
\end{abstract}

\preprint{OCU-PHYS-434}
\preprint{AP-GR-128}


\date{\today}
\maketitle

\section{Introduction}\label{intro}
Higher-dimensional spacetimes are popular subjects in fundamental theoretical physics, 
mainly in the context of unified theories of interactions. 
In order to compromise the higher dimensions with the experimentally observable four dimensions, 
the spacetimes would be decomposed into four-dimensional conventional spacetime in a large size 
and compactified extra dimensions in a small size 
(see ref.~\cite{Appelquist:1987nr}, for example, and references there in). 
We call the higher-dimensional spacetimes with such structures Kaluza-Klein spacetimes. 
The three dimensions has cosmological size while the extra dimensions should be tiny enough not 
to observe by experiments in laboratories. 
Why the large discrepancy of the sizes appears? 
It would be a natural idea that the extra dimensions shrink by evolution of the 
universe~\cite{Chodos:1979vk, Freund:1982pg, RandjbarDaemi:1983jz}. 
If the inflation of the universe occurs by the contraction of the 
extra dimensions~\cite{Sahdev:1988fp, Ishihara:1984wx, Ishihara:1986if}, 
a huge difference of the sizes can emerge.

Recently, black holes in higher dimensions gather much attention because 
they have rich variety in contrast to four-dimensional cases. 
Extended black objects, black strings for example, are found in the higher dimensions~\cite{
Horowitz:1991cd, Gibbons:1994vm, Bleyer:1994wr, Horowitz:2002ym}.
It is striking that exact solutions with horizon topology $S^2 \times S^1$ are found~\cite{Emparan:2001wn}
in addition to rotating black holes with spherical horizon topology~\cite{Myers:1986un} 
in asymptotically flat spacetimes. 
After this discovery, lots of related black objects 
were found (see ref.~\cite{Emparan:2008eg}, as a review). 
It would be also important to investigate black hole solutions 
asymptote to the Kaluza-Klein spacetimes. 
Exact solutions which asymptote to a locally flat spacetime 
with a twisted $S^1$ extra dimension were constructed~\cite{Dobiasch:1981vh, Gibbons:1985ac,
Ishihara:2005dp, Nakagawa:2008rm,  Tomizawa:2008hw, 
Matsuno:2012hf,  Tomizawa:2012nk, Stelea:2012ph}.

There is a class of exact solutions to the Einstein-Maxwell equations constructed 
by using harmonic functions on Ricci flat base spaces with Euclidean signature, 
where the metric functions and the gauge field are described by various kinds of 
the harmonics. 
If we take the harmonic function for a point source properly, 
the solutions describe extremely charged black holes. 
It is straight forward to construct multi-black hole solutions by superposition of the 
harmonic functions. 
A pioneering work was done by Majumdar and Papapetrou~\cite{Majumdar:1947eu, Papaetrou:1947ib} in four 
dimensions,
and higher-dimensional solutions of this class are constructed as a variety 
of black holes~\cite{Myers:1986rx, Breckenridge:1996is, Gauntlett:2002nw, Gaiotto:2005gf, 
Ishihara:2006iv,Ishihara:2006pb,  Matsuno:2008fn, Tatsuoka:2011tx, Matsuno:2012ge, Matsuno:2015ega}. 
In the case of hyper-K\"ahler base space, the solutions are supersymmetric. 
As generalizations of four-dimensional time-dependent multi-black hole 
solutions~\cite{Kastor:1992nn, Chimento:2012mg, Chimento:2014afa},
higher-dimensional solutions including a cosmological constant~\cite{
London:1995ib,Klemm:2000vn, Ishihara:2006ig, Ida:2007vi, Matsuno:2007ts,  Kimura:2009er}
and without cosmological constant~\cite{Gibbons:2005rt, Maeda:2009zi, Maeda:2009ds, Maeda:2010ja, Kanou:2014rya} 
are obtained  by using harmonics.

Indeed, exact solutions are constructed by solving the Laplace equation 
on a Ricci flat base space, 
but the solutions do not always describe black objects. 
If we want to construct black hole solutions by using harmonics for a point source 
on the Euclidean Taub-NUT 
space, the point source should be on the NUT singularity for the appearance 
of horizon~\cite{Ishihara:2006iv}. 
By harmonics for a line source on the four-dimensional flat Euclidean space
a naked singularity appears.

In the present paper, we construct an exact solution combining a Kaluza-Klein 
cosmological solution 
and a black string solution by using a harmonic function. 
The solution describes a charged black string in a five-dimensional Kasner universe where 
three-dimensional space expands while one dimension contracts as time increases.

In the next section, the metric and the gauge field are presented,  
and it is shown that the spacetime has a regular event horizon where analytical extension is possible. 
Although the spacetime is dynamical, {\it i.e.}, it has no timelike Killing vector field, 
the geometry near the horizon becomes static quickly in the late time, 
then the size of horizon does not change. 
This property is characterized by the existence of asymptotic Killing generator near the horizon.

In section {I\!I\!I}, we study geodesic motions of timelike and null test particles. 
We show the existence of quasi stable circular orbits of massive particle around the 
black string. The radius of the circular orbit decreases gradually, and finally falls into the 
horizon. Analogous to the four-dimensional Schwarzschild black hole, quasi innermost stable 
circular orbit appears. There also exists unstable circular orbits for massless particles.

In section {I\!V}, we discuss whether the Gregory-Laflamme instability occurs or not briefly. 
The section {V} is devoted to summary and discussion.

\section{Black strings in Kaluza-Klein universe}\label{sec:solution}
\subsection{Solution}

We consider time-dependent charged black strings 
which are exact solutions of the five-dimensional Einstein-Maxwell theory with the action
\begin{align}
 S = \frac{1}{16\pi} \int d^5 x \sqrt{-g} 
  \left( R - F_{\mu\nu} F^{\mu\nu} \right) .
\end{align}
The metric and the Maxwell field are given by
\begin{align}
 ds^2 &= -H^{-2} dt^2 + H \left[ \frac{ t }{ t_0} \left( dr^2 + r^2 d\Omega^2_{\rm S^2} \right) 
+ \frac{ t _0 }{ t } dw ^2 \right] , 
\label{eq:metric3}
\\
A_\mu dx^\mu &= \pm \frac{ \sqrt{3} }{ 2 } H^{-1} dt, 
\label{eq:field3}
\end{align}
where the function $H$ is given by 
\begin{align}
	H = 1 + \frac{ M }{ r } , 
\label{eq:h3}
\end{align}
$d\Omega^2_{\rm S^2} = d\theta^2 + \sin ^2 \theta d\phi^2$ is the metric of 
unit two-dimensional sphere, S$^2$, 
$t_0$ and $M$ are non-negative constants.\footnote{
Taking the limit $N \to 0$ appears in the black hole solution in a Kaluza-Klein universe discussed 
in ref.~\cite{Kanou:2014rya}, 
one can obtain the metric~\eqref{eq:metric3} and the Maxwell field~\eqref{eq:field3}. 
We generalize the solution~\eqref{eq:metric3} to multi-black string and multi-black hole solutions 
in Appendix~\ref{extensionmulti}.}   
For the spacetime signature being $(- , + , + , + , +)$, we should require the inequality $H t / t_ 0 > 0$.

In the limit $M \to 0$ or $r \to + \infty$ with $t = {\rm finite}$, 
the field strength of the Maxwell field~\eqref{eq:field3} vanishes, and 
the metric~\eqref{eq:metric3} reduces to that of the five-dimensional vacuum Kasner 
universe,  
\begin{align}
ds^2 = - dt^2 + \frac{t}{t_0} \left( dr^2 + r^2 d\Omega^2_{\rm S^2} \right) 
+ \frac{t_0}{t} dw ^2 , 
\label{eq:kasner}
\end{align}
which describes time evolution of an anisotropic and homogeneous universe.  
If the extra dimension labeled by the coordinate $w$ is compactified by periodic identification, 
its size becomes small enough after the evolution of the universe. 
The metric~\eqref{eq:metric3}, as same as the metric~\eqref{eq:kasner}, 
has a null infinity at $t = +\infty ,~ r = +\infty$ with $r/t= {\rm finite}$.

The Kretschmann scalar is given by
\begin{align}
R^{\mu\nu\rho\sigma}R_{\mu\nu\rho\sigma} 
= \frac{18 (r+M)^5 \left[ (r+M)^5 - t_0 M^2 t r \right] 
+ \left(t_0 M t r \right)^2 (144 r^2 + 48 M r + 31 M^2)}
{4 t^4 r^4 (r+M) ^6}.
\end{align}
The singularity $t = 0$ with $r = {\rm const.}$ corresponds to the initial 
cosmological singularity for the Kasner universe.
Since the norm of the normal vector to a $t = {\rm const.}$ surface
$g^{\mu \nu}(dt)_\mu(dt)_\nu = - H^2$
is negative, the curvature singularity $t = 0$ is spacelike. 
We should pay attention to a limiting surface $r \to 0$ and $t \to \infty$ keeping $rt={\rm const.}$, 
because the Kretschmann scalar is finite. The sizes of the two-dimensional sphere and the extra dimension 
become finite at the limit. 
As will be shown later, we find the surface is the event horizon and we extend the metric~\eqref{eq:metric3} 
across the surface.\footnote{As far as we consider analytic extension across the event horizon, we do not need to care about $r = -M$ singularity.}

\subsection{Analytic extension across the event horizon}
Instead of the original coordinates in~\eqref{eq:metric3} which do not cover $r=0$, 
we construct coordinates covering 
the surface $r = 0,~ t = \infty$ with $r t = {\rm const.}$ 
using a set of null geodesics.  
We investigate the possibility of extension by using null geodesics starting 
from the outer region $r>0$.

If we restrict our attention to the null geodesics 
confined in the $t$-$r$ plane, 
the null geodesics are determined by the null condition, 
\begin{align}
-H^{-2} dt^2 + H \frac{t}{t_0} dr^2 = 0, 
\end{align} 
namely
\begin{align} 
\left( \frac{dt}{dr} \right)^2 = \frac{t}{t_0} \left( \frac{M}{r} + 1 \right) ^3 .
\label{eq:nullcond2}
\end{align} 
We use an approximate solution in the form 
\begin{align}
	t r = \frac{\left( 2 M^2 + u \sqrt{M r}- 3 M r \right)^2}{4 M t_0},
\label{eq:nullcondsolassume} 
\end{align}
where $u$ is an arbitrary parameter. 
The curves~\eqref{eq:nullcondsolassume} 
are approximately ingoing future null geodesics in the vicinity of $r = 0$ 
that attain the coordinate boundary.  
The free parameter $u$, which labels the curves, can be used as a new coordinate.\footnote{
Although we can solve the equation~\eqref{eq:nullcond2} analytically as 
\begin{align} 
u = 2 \sqrt{t_0 t} + (r - 2 M)\sqrt{1 + \frac{M}{r}}
+ 3 M{\rm arcsinh} \left(\sqrt{\frac{r}{M}} \right),
\notag
\end{align}
it is convenient to use the approximate solution~\eqref{eq:nullcondsolassume}.
Eq.\eqref{eq:nullcondsolassume} coincides with this analytic solution up to ${\cal O}(r^{3/2})$.
}

Now, we introduce new coordinates $(u, \rho)$ as 
\begin{align}
r &= \rho^2/M ,
\label{urhocoord0}
\\ 
t &= 
\frac{\left( 2 M^2 + (u - 3 \rho) \rho \right)^2}{4 t_0 \rho^2},
\label{urhocoord}
\end{align} 
then we rewrite the metric~\eqref{eq:metric3} and the Maxwell field~\eqref{eq:field3} 
in the $(u, \rho)$ coordinates as 
\begin{align} 
ds^2 &= 
\left(
\frac{2 M^2 + (u - 3 \rho)\rho}{2 t_0 (M^2 + \rho^2)}
\right)^2
\bigg[
-\rho^2 du^2 + 2 (2 M^2 + 3 \rho^2) du d\rho
+
\rho^2 \left(3 + \frac{4\rho^2}{M^2}\right) d\rho^2
\notag\\
& \quad + 
\frac{(M^2 + \rho^2)^3}{M^2} d\Omega_{\rm S ^2}^2
+
\frac{16 t_0^4 (M^2 + \rho^2)^3}{(2 M^2 + (u - 3 \rho)\rho)^4}
dw^2
\bigg],
\label{eq:metricanalytic}
\\
A_\mu dx^\mu &= 
\pm \frac{\sqrt{3}  \rho }{4 t_0 (M^2 + \rho^2)}
\Big[
\big(4 M^2 + (u-\rho)\rho\big) du
- u \rho d\rho
\Big], 
\label{eq:fieldanalytic}
\end{align} 
where we omit a pure gauge term in $A_\mu dx^\mu$. 
The metric and the Maxwell field are regular at $\rho = 0$.  
In the limit $\rho \to 0$ with $u = {\rm finite}$  
(equivalently, $r \to 0 ,~ t \to \infty$ with $t r =  M^3 / t_0$), 
the metric~\eqref{eq:metricanalytic}
 behaves as 
\begin{align} 
ds^2 &\to \frac{4 M^2}{t_0^2} du d\rho + \frac{M^4}{t_0 ^2} d\Omega_{\rm S ^2}^2 + \frac{t_0^2}{M^2} dw^2.
\label{eq:metriclimit}
\end{align}
We also see that the $\rho = 0$ surface is a null hypersurface, 
and the angular part of the metric, 
which describes $\rm S ^2 \times \rm R ^1$, 
does not depend on time. 
Since all the metric components in~\eqref{eq:metricanalytic} are 
analytic function of $\rho$ at $\rho = 0$,
the spacetime with the metric~\eqref{eq:metricanalytic}
gives an analytic extension of the original spacetime~\eqref{eq:metric3}.
We can see that 
the inner region $\rho < 0$ is a time reversal of the outer region $\rho > 0$ 
since the metric~\eqref{eq:metricanalytic} is invariant under the transformation 
\begin{align}
\rho \to -\rho, \quad u \to -u . 
\end{align}

The outer region $\rho > 0$ 
becomes asymptotically the Kasner universe 
described by~\eqref{eq:kasner}, 
then it has a future null infinity. 
However, any null geodesic starting from a point in 
the inner region 
cannot reach the future null infinity. Therefore, the $\rho = 0$ surface is an event horizon. 
The exact solution~\eqref{eq:metric3} with~\eqref{eq:field3} indeed represents 
the charged black string in the five-dimensional 
anisotropically 
expanding Kaluza-Klein universe.

\begin{figure}[h]
\begin{center}
 \includegraphics[width=0.7\linewidth,clip]{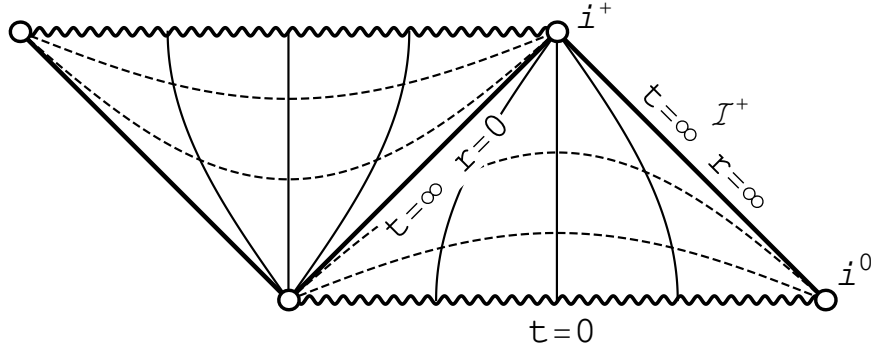}
\end{center}
\caption{Penrose diagram of $t$-$r$ plane.
The outer region and the inner region are joined at
the event horizon, $r=0$ and $t=\infty$ with $rt=M^3/t_0$.
The null infinity exists at $r=+\infty$ and $t=+\infty$.
The wavy lines are curvature singularities.
Dashed curves denote $t ={\rm const.}$ surfaces, and
thin solid curves denote $r ={\rm const.}$ surfaces.}
\label{fig:penrose}
\end{figure}

The Penrose diagram of the spacetime with the metric~\eqref{eq:metric3} is shown in Fig.~\ref{fig:penrose}.
In the outer region,  
the geometry looks like the five-dimensional Kasner universe described by~\eqref{eq:kasner} 
in the far region, $r \gg M$, where the three-dimensional space expands,
while the compact extra dimension shrinks with the time evolution.  
There are a null infinity at $r = \infty ,~ t = \infty$
and spacelike singularity at $t = 0$.
A null hypersurface 
$r \to 0, t \to \infty$ with $rt=M^3/t_0$ is the event horizon.
The inner region of the event horizon is the time reversal of the outer region,
and these two regions are attached with each other at the event horizon $\rho = 0$.

\subsection{Approximate staticity near the event horizon}
\label{staticity}

The spacetime is dynamical because the metric~\eqref{eq:metric3} 
does not admit any timelike Killing vector. 
However, as shown in~\eqref{eq:metriclimit},
since the metric on the horizon does not depend on time, 
it is suggested that the metric is approximately static near the horizon. 
In fact, the event horizon is an example of the \lq asymptotic Killing horizon\rq\ 
defined in~\cite{Koga:2001vq, Koga:2006ez}.  
In this paper, we consider an $n$-th order asymptotic Killing horizon
as a generalization. 
\\

{\it Definition.}~~{\it
A null hypersurface ${\cal H}$ is an $n$-th order asymptotic Killing horizon 
if there exist a scalar function $\Phi$ and a vector field $\xi^\mu$, such that
(i)~$\Phi =0, d\Phi \neq 0$ on ${\cal H}$,
(ii)~${\cal L}_{\xi^\mu} g_{\alpha \beta} = {\cal O}(\Phi^n)$, 
$\xi^\mu \xi_\mu = {\cal O}(\Phi)$, 
$\xi^\mu \nabla_\mu \Phi = {\cal O}(\Phi)$, 
where ${\cal L_{\xi^\mu}}$ denotes the Lie derivative with respect to a 
vector field $\xi^\mu$.
}
\\

The first equation of the condition (ii) is Killing equation up to the $n$-th order of $\Phi$,
and the rest of the condition (ii) requires $\xi^\mu$ is a generator of the horizon.
We call a solution $\xi^\mu_{[n{\rm th}]}$ $n$-th order asymptotic Killing generator.\footnote{
Note that $n = 1$ case is discussed in Refs.~\cite{Koga:2001vq, Koga:2006ez}.}
By choosing $\Phi = \rho$ in $(u,\rho)$ coordinate, we find a first order 
asymptotic Killing generator
\begin{align}
\xi^\mu_{[1{\rm st}]} &= 
f \left(\frac{\partial}{\partial u}\right)^\mu
- 
\rho
\partial_u f
\left(\frac{\partial}{\partial \rho}\right)^\mu
-
\frac{2\rho \partial_\theta f}{M^2}
\left(\frac{\partial}{\partial \theta}\right)^\mu
-
\frac{2\rho \partial_\phi f}{M^2 \sin^2\theta }
\left(\frac{\partial}{\partial \phi}\right)^\mu
\notag\\&\qquad
-
\frac{2 M^4 \rho \partial_w f}{t_0^4}
\left(\frac{\partial}{\partial w}\right)^\mu
+
{\cal O}(\rho^2),
\label{1stasymptotickilling}
\end{align}
where $f$ is an arbitrary function of $u,\theta,\phi,w$.
We should note that this functional degrees of freedom generally appears in the 
first order asymptotic Killing horizon~\cite{Koga:2001vq, Koga:2006ez}.
Namely, the first order asymptotic Killing generator is not determined uniquely by the geometry.

The present metric~\eqref{eq:metricanalytic} also admits a second order asymptotic Killing generator
\begin{align}
\xi^\mu_{[2{\rm nd}]} &= 
C
\left[
u
\left(\frac{\partial}{\partial u}\right)^\mu
- 
\rho
\left(\frac{\partial}{\partial \rho}\right)^\mu
\right]
+
{\cal O}(\rho^3),
\end{align}
where $C$ is an arbitrary constant.
This is the unique second order asymptotic Killing generator. 
We understand that the second order asymptotic Killing generator $\xi^\mu_{[2{\rm nd}]}$ 
characterizes approximate staticity of the geometry near the event horizon.\footnote{
From the facts that
the spacetime admits second order asymptotic Killing generator
and the Maxwell field satisfies energy conditions,
we can also show that the horizon is an isolated horizon~\cite{Ashtekar:2004cn}.
}
Since we can show that $\xi^\mu_{[2{\rm nd}]}$ is 
timelike in $\rho u > 0$ and spacelike in $\rho u < 0$ 
near the horizon, 
the spacetime has second order {\it timelike} asymptotic Killing generator only in the region 
$u > 0$ outside the horizon.\footnote{
Note that $\xi^\mu_{[2{\rm nd}]}$ has a fixed point at $u = 0, \rho = 0$ 
like a bifurcation point of a usual static black hole.}
By choosing $\xi^\mu_{[2{\rm nd}]}$ as a time coordinate basis,
we can find an approximately static coordinate $(\bar{T}, \bar{R})$ in the region $u > 0$ as
\begin{align}
u &= e^{\bar{T}/M} (\bar{R} + M),~~
\rho = e^{-\bar{T}/M} \frac{M \bar{R}}{(\bar{R} + M)},
\label{barcoordinate}
\end{align}
where 
$\partial_{\bar T} = M^{-1} (u \partial_u - \rho \partial_\rho)$.
In this coordinate, the metric becomes
\begin{align}
ds^2 & =
- \frac{\bar{R}(2M + \bar{R})^2(4M +\bar{R})}{4 M^2 t_0^2} d\bar{T}^2
+
\frac{(2 M+\bar{R})^2(2 M^2 - 2 M \bar{R} -\bar{R}^2 )}{2 M(M+\bar{R}) t_0^2}
d\bar{T} d\bar{R}
\notag\\ & \qquad 
+
\frac{(2 M - \bar{R})(2 M+\bar{R})^3}{4 (M+\bar{R})^2 t_0^2}
d\bar{R}^2
+
\frac{M^2(2 M + \bar{R})^2}{4t_0^2}
d\Omega_{\rm S^2}^2 
\notag\\ & \qquad  
+ 
\frac{4 t_0^2}{(2M + \bar{R})^2} dw^2
+
{\cal O}\left( e^{-2\bar{T}/M} \bar{R}^2 / M^2 \right).
\label{metricbartr}
\end{align}
We can see that the geometry near event horizon is static at the order of ${\cal O}(\bar{R}/M)$ as expected.

Furthermore, we can also see that the metric near the horizon 
rapidly approaches the time independent form 
in Eq.~\eqref{metricbartr}
at late time $e^{- 2 \bar{T}/M} \ll 1$.
In fact, the time independent part 
in Eq.~\eqref{metricbartr} can be written in a simple form 
\begin{align}
 ds^2_{\rm TI} &= - \frac{R ^2 \left( R ^2 - R_h ^2 \right)}{R_h^4} dT^2 
+ \frac{4 R ^2}{R ^2 - R_h ^2} dR ^2 
+ R ^2 d\Omega^2_{\rm S^2} 
+ \frac{R_h^2}{R ^2} dW ^2 ,
\label{eq:metric4}
\end{align}
by introducing coordinates $(T,R,W)$ as
\begin{align}
\bar{T} = \frac{1}{2} \sqrt{\frac{t_0}{R_h}} T + \frac{\sqrt{t_0 R_h}}{2} \ln \frac{R- R_h}{(R + R_h)(2R-R_h)^2},~
\bar{R} =  \frac{2\sqrt{t_0}(R - R_h)}{\sqrt{R_h}},~
w = \sqrt{\frac{R_h}{ t_0}}~W
\end{align}
with $R_h := M^2/t_0$.
The metric~\eqref{eq:metric4} is a limiting case of static charged black string solutions derived in 
ref.~\cite{Bleyer:1994wr} (see Appendix~\ref{appendix:hmblackstring}).\footnote{
Note that we can also derive Eq.~\eqref{eq:metric4} by taking 
a limit $t \to t/\epsilon, r \to \epsilon r$ and $\epsilon \to 0$ for the metric~\eqref{eq:metric3}.
In this limit the metric has the form of
\begin{align}
 ds^2 &= - \frac{r^2}{M^2} dt^2 
+ \frac{t M}{t_0 r} dr ^2 
+ \frac{t r M}{t_0} d\Omega^2_{\rm S^2} 
+ \frac{t_0 M}{t r} dw^2.
\end{align}
Though there is an apparent time dependence,
by introducing coordinates
\begin{align}
R^2 = \frac{M t r}{t_0},~
dT = \frac{M^2}{t_0 t } dt + \frac{2 M^6/t_0^3}{R \left( R ^2 - M^4/t_0^2 \right)} dR,~
W = \frac{t_0}{M} w,
\label{coordRTW}
\end{align}
we obtain the metric form (\ref{eq:metric4}).
}

Since the time dependence appears from ${\cal O}(\bar{R}^2/M^2)$, we can expect the curvature tensor 
also has a time dependence on the horizon.
While the Einstein tensor and the Kretschmann scalar take constant on the horizon, 
Weyl tensor has time dependence there. The leading term is same as Weyl tensor of 
Eq.~\eqref{eq:metric4} and the time dependence decays ${\cal O}(e^{-2\bar{T}/M})$ at late time.

\subsection{Expansion of a null geodesic congruence}

We calculate the expansions of the null vector fields 
emanating from a closed surface $r = {\rm const.}$ on a $t = {\rm const.}$ slice. 
The expansions are defined by 
\begin{eqnarray}
 \theta ^ \pm = h^{\mu \nu} \nabla_\mu k ^{(\pm)} _\nu ,
\end{eqnarray}
where $k ^{(\pm) \mu}$
denote future null vector fields, and 
$h_{\mu \nu} = g_{\mu \nu} + k^{(+)} _\mu k^{(-)} _\nu + k^{(-)} _\mu k^{(+)} _\nu$. 
We choose $k ^{(\pm) \mu}$
as 
\begin{eqnarray}
 k ^{(+) \mu} \frac{\partial}{\partial x^\mu} 
 &=& \sqrt{\frac{t}{2 t_0}} \frac{\partial}{\partial t} + \frac{1}{H \sqrt{2 H}} \frac{\partial}{\partial r} ,
\label{nullveco}
\\ 
 k ^{(-) \mu} \frac{\partial}{\partial x^\mu} 
 &=& H^2 \sqrt{\frac{t_0}{2 t}} \frac{\partial}{\partial t} - \frac{t_0}{t} \sqrt{\frac{H}{2}} \frac{\partial}{\partial r} ,
\label{nullveci}
\end{eqnarray}
such that the direction of the vector fields are to be 
increasing $r$ coordinates for $(+)$, and decreasing for $(-)$. 
Since $k^{(\pm) \mu}$ satisfy the relations $k^{(-) \mu} \nabla_\mu k^{(-) \nu} = 0$ and 
$g_{\mu \nu} k^{(+) \mu} k^{(-) \nu} = -1$, they are regular everywhere in the spacetime.
Note that $k^{(\pm) \mu}$ on and inside the event horizon should be understood as
analytic extensions of them across the horizon.
The metric on ${\rm S}^2 \times {\rm R}^1$ ($r ={\rm const.}$, $t = {\rm const.}$), 
$h_{\mu \nu}$, becomes 
\begin{eqnarray}
 h_{\mu \nu} dx^\mu dx ^\nu = H \frac{t}{t_0} r^2 d\Omega ^2 _{\rm S ^2} + H \frac{t_0}{t} dw^2 .
\label{hmunu}
\end{eqnarray}
The expansions of the null geodesic congruences on the three-dimensional space~\eqref{hmunu} 
are obtained as
\begin{eqnarray}
 \theta ^+ &=& \frac{r^3 H^3 + r (4r+M) \sqrt{t_0 t H} }{2 r^3 H^3 \sqrt{2 t_0 t} } ,
\\
 \theta ^- &=& \frac{r^3 H^3 - r (4r+M) \sqrt{t_0 t H} }{2 t r^3 H \sqrt{2 t / t_0} }  .
\end{eqnarray}

\begin{figure}[htbp]
\begin{center}
 \includegraphics[width=0.7\linewidth]{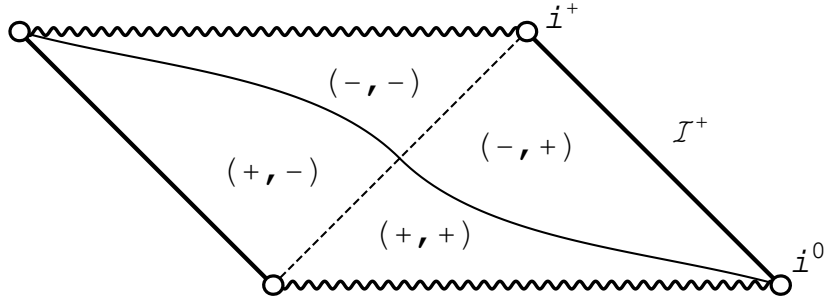}
\end{center}
\caption{
The sign of $\theta^\pm$ and $\theta ^\pm = 0$ surfaces
in the Penrose diagram.
Pairs of $(\pm , \pm)$ denote the sign of $(\theta^- , \theta^+)$.
A broken line and a solid curve denote the $\theta^+ = 0$ and $\theta^- = 0$ surfaces,
respectively.
}
\label{fig:penrose_expansions}
\end{figure}

We show the sign of $\theta ^\pm$ and $\theta ^\pm = 0$ surfaces in Fig.~\ref{fig:penrose_expansions}. 
Since $\theta ^+ = 0$ in the limit $r \to 0 ,~ t \to \infty$ with $t r = M^3 / t_0$,  
we see that, though the spacetime is dynamical, 
the event horizon, $r = 0 ,~ t = \infty$ with $t r = M^3 / t_0$ surface, is an apparent horizon.
We also see that outside the black string, $r > 0$, 
$\theta ^+$ is positive,   
and a region $(\theta ^- , \theta ^+) = (+ , +)$ appears near the initial singularity like an expanding universe. 
Inside the black string, there is a trapped region, $(\theta ^- , \theta ^+) = (- , -)$, 
like a static black hole spacetime.

Using $(u, \rho)$ coordinates~\eqref{urhocoord0} and~\eqref{urhocoord} which cover the event horizon,
we can easily see the behavior of $\theta^\pm$ near the horizon as 
$\theta^+ = \rho/(\sqrt{2}M^2) + {\cal O}(\rho^2)$ and   
$\theta^- = -u t_0 ^2/(4\sqrt{2}M^4) + {\cal O}(\rho)$.
The sign of $\theta^+$ changes on $\rho = 0$ surface, 
and the sign of $\theta^-$ does on the $u = 0$ surface near the horizon. 
Both expansions become zero at the point $(u, \rho) = (0,0)$ which is a fixed point of
$\xi^\mu_{[2{\rm nd}]}$ on the horizon.

\section{Motion of a test particle}
A test particle provides us geometrical information of a spacetime as a probe, 
then we study motion of it in the present spacetime.  
Motions of a test particle in the metric~\eqref{eq:metric3} is 
governed by the Lagrangian
\begin{align}
\mathcal L = \frac12\left[-H(r)^{-2} \dot t^2 
+ H(r) a(t)^2 \left( \dot r^2 
+  r^2 \dot \theta^2 +  r^2 \sin ^2 \theta \dot \phi^2 \right)
+ H(r)a(t)^{-2} \dot w ^2\right] ,
\label{lagrangian}
\end{align}
where dot denotes derivative with respect to the proper time of the particle, 
and the function $a(t)^2=t/t_0$ plays the role of cosmological scale factor for 
the expanding three dimensions. 
The particle has conserved quantities 
\begin{align}
L = H a^2 r^2 \sin^2 \theta \dot \phi,
\quad {\rm and}\quad
p_w = H a^{-2} \dot w .
\end{align}
Since the spacetime has the spherical symmetry, we restrict that the particle moves in the plane 
$\theta=\pi/2$ without loss of generality.

The Euler-Lagrange equations for~\eqref{lagrangian} with $\theta=\pi/2$ yield 
\begin{align}
& -\frac{d}{d\tau}(H^{-2} \dot t)
= a\frac{da}{dt} \left(H\dot r^2 + \frac{1}{H a^4 r^2} L^2 - H^{-1}p_w^2 \right),
\cr
&  \frac{d}{d\tau} \left(H a^2 \dot r\right)
= H^{-3}\frac{dH}{dr} \dot t^2 
+\frac12\frac{dH}{dr}\left( a^2\dot r^2 + \frac{L^2}{H^2a^2r^2} \right)
+ \frac{L^2}{Ha^2r^3} 
+ \frac12\frac{dH}{dr}\frac{a^2 p_w^2}{H^2}. 
\label{E-L_eqs}
\end{align}

For a massive particle, we have
\begin{align}
 g_{\mu\nu}\dot x^\mu \dot x^\nu
= - H^{-2} \dot t^2 + H a^2 \dot r ^2 + \frac{L^2}{H a^2 r^2}+\frac{a^2}{H}p_w^2  =-1. 
\label{normalization}
\end{align}

\subsection{Quasi circular orbits}

We should note that the canonical momentum conjugate to the time $t$, 
\begin{align}
	p_t=\frac{\partial\mathcal L}{\partial \dot t}=-H^{-2} \dot t,  
\end{align}
is not conserved since the metric depends on $t$ through the function $a(t)$. 
This is because the metric describes a black string residing in an evolving universe. 
In a much late stage of the universe, say $t=t_1$, the time scale of the evolution, 
\begin{align}
	a(t)\left(\frac{da(t)}{dt}\right)^{-1}= 2 t_1 
\end{align}
becomes much longer than the time scale of an orbiting particle near the black string. 
For this particle, $a(t)$ is almost constant at $t\sim t_1$ during its typical motion, 
then $p_t$ is almost constant, say $-E$.  
In this case, \eqref{normalization} is rewritten as 
\begin{align}
& H^{-1} a^2 \dot r ^2 + U_{\rm eff} = E^2 ,
\end{align}
where 
\begin{align}
 U_{\rm eff} = H^{-2} \left( 1 +\frac{L^2}{H a^2 r^2} +\frac{a^2}{H}p_w^2 \right). 
\label{U_eff}
\end{align}
then, we can analyze orbits of the particle using the effective potential~\eqref{U_eff} 
with a constant $a(t_1)$.

Here, taking the cosmological scale factor into account, we introduce a radial coordinate 
that denotes \lq physical length\rq\ on a time-slice $t=t_1$ as 
\begin{align}
	\tilde r =a(t_1)~r. 
\end{align} 
Using $\tilde r$, we represent the effective potential~\eqref{U_eff} as 
\begin{align}
	U_{\rm eff}(\tilde r, t_1) 
		= \frac{\tilde r  L^2+\tilde r^3 a(t_1)^2 p_w^2 +\tilde r^2 (\tilde M+\tilde r) }
			{(\tilde M+\tilde r)^3}, 
\label{U_eff_bar}
\end{align}
where
\begin{align}
	\tilde M=a(t_1)M. 
\end{align}

\begin{figure}[h]
\begin{center}
 \includegraphics[width=10cm,clip]{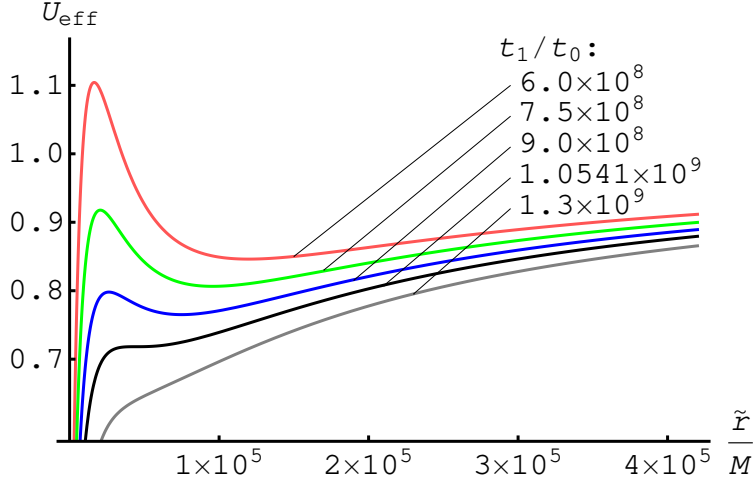}
\end{center}
\caption{
The behavior of the effective potential~\eqref{U_eff_3dim} versus $\tilde r / M$
in various $t_1 / t_0$ for $L / M = 62722$.
The effective potential evolves as $t_1 / t_0$ increases: 
$t_1 / t_0 = 6 \times 10^8$ (red curve),
$t_1 / t_0 = 7.5 \times 10^8$ (green curve),
$t_1 / t_0 = 9 \times 10^8$ (blue curve),
$t_1 / t_0 = (2 - \sqrt 3) L^2 / M^2 \simeq 1.0541 \times 10^9$ (black curve), and
$t_1 / t_0 = 1.3 \times 10^9$ (gray curve).  
}
\label{fig:U_eff}
\end{figure}

The size of compactified dimension $w$ becomes 
very small owing to the evolution of the universe. 
It would be expected that the momentum $p_w$ conjugate to $w$ is hardly excited \cite{Ishihara:1985zm}. 
Then, we consider the case $p_w=0$. 
The restriction of the metric~\eqref{eq:metric3} and the particle motion on $w={\rm const.}$ 
surface would be a model of a four-dimensional black hole in an expanding universe. 
In this case, \eqref{U_eff_bar} reduces to 
\begin{align}
	&U_{\rm eff}(\tilde r, t_1) 
		= \frac{\tilde r L^2+\tilde r^2 (\tilde M+\tilde r)}
			{(\tilde M+\tilde r)^3}. 
\label{U_eff_3dim}
\end{align}
The effective potential \eqref{U_eff_3dim} is shown as the function of $\tilde r$ in Fig.~\ref{fig:U_eff}. 
Since $\tilde M$ depends on time $t_1$, the effective potential~\eqref{U_eff_3dim} gradually changes 
in its shape as the universe expands.

The existence of an innermost stable circular orbit (ISCO) is a feature of four-dimensional 
stationary black holes.  
It is an interesting question whether the ISCO exists or not in the present case. 
Then, we concentrate on circular orbits.

If $a(t_1)$ and $E$ were constants, for a fixed $L$, 
there exist a stable circular orbit and an unstable one specified by 
\begin{align}
	U_{\rm eff}-E^2=0 ,\quad \mbox{and}\quad \frac{dU_{\rm eff}}{d\tilde r} = 0. 
\label{cond_for_circle} 
\end{align}
In fact, since $a(t_1)$ and $E$ gradually changed as time increases, 
the radius of the orbits specified by~\eqref{cond_for_circle} 
are gradually changed. Then, we call these orbits \lq quasi circular orbits\rq. 
The radius of the stable quasi circular orbit is
\begin{align}
	\tilde r_s= \frac{1}{2\tilde M}\left( L^2-\tilde M^2+ \sqrt{(L^2-\tilde M^2)^2-2\tilde M^2L^2} \right),
\end{align}
and the radius of the unstable quasi circular orbit is
\begin{align}
	\tilde r_u= \frac{1}{2\tilde M}\left( L^2-\tilde M^2- \sqrt{(L^2-\tilde M^2)^2-2\tilde M^2L^2} \right).
\end{align}
In order to have real positive $\tilde r_s$, $L^2 \geq (2+\sqrt{3})\tilde M^2$. 
In the case of $L^2 = (2+\sqrt{3})\tilde M^2$, we have the radius of the quasi innermost stable 
circular orbit (qISCO) is given by
\begin{align}
 \tilde r_{\rm qISCO}=\frac{1+\sqrt{3}}{2}\tilde M, 
\end{align}
and a qISCO particle has the angular momentum 
\begin{align}
	L=\pm L_{\rm qISCO}=\pm \frac{1+\sqrt{3}}{\sqrt{2}}\tilde M. 
\end{align}
The ratio $r/M$ for the qISCO is not smaller than unity, then there does not exist 
quasi circular orbit in the approximately static region near the horizon.

\begin{figure}[h]
\begin{center}
 \includegraphics[width=10cm,clip]{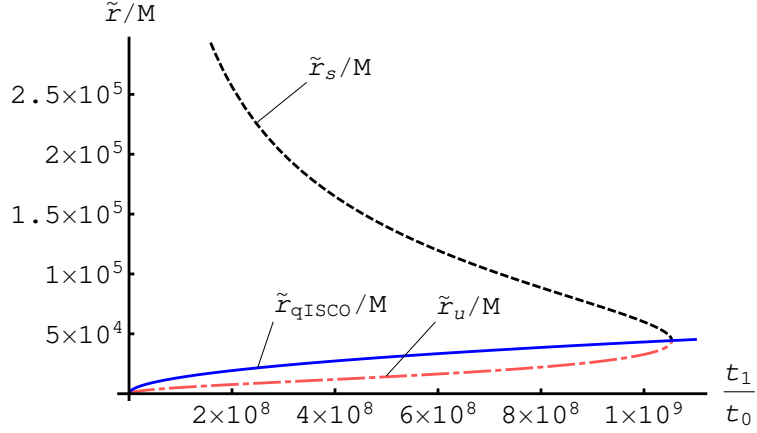}
\end{center}
\caption{
The behavior of the radius of the quasi circular orbits,
$\tilde r _s / M$ (black broken curve), $\tilde r _u / M$ (red dashed-dotted curve),
and $\tilde r _{\rm qISCO} / M$ (blue solid curve), 
versus $t_1 / t_0$ for $L / M = 62722$.
At the time $t_1/t_0=(2-\sqrt{3})L^2/M^2 \simeq 1.0541 \times 10^9$, 
$\tilde r_s$ and $\tilde r_u$  merge together at $\tilde r_{\rm qISCO}$.
}
\label{r_of_t}
\end{figure}

Note the time dependence of $\tilde M$, we see that 
$\tilde r_s$, $\tilde r_u$, and $\tilde r_{\rm qISCO}$ are time dependent.
Suppose a quasi circular particle at $\tilde r_s (> \tilde r_{\rm qISCO})$ for a fixed angular momentum $L$ at a time $t_1$ 
in the present metric with fixed parameters $M$ and $t_0$.  
As the time $t_1$ increases the radius $\tilde r_s$ decreases and the radius $\tilde r_u$ increases (see Fig.~\ref{r_of_t}). 
At the time $t_1/t_0=(2-\sqrt{3})L^2/M^2$, $\tilde r_s$ and $\tilde r_u$  merge 
together at $\tilde r_{\rm qISCO}$. 
The particle with $L$ has no circular orbit after that time, 
then the particle plunges into the black string.

\begin{figure}[h]
\begin{center}
 \includegraphics[width=10cm,clip]{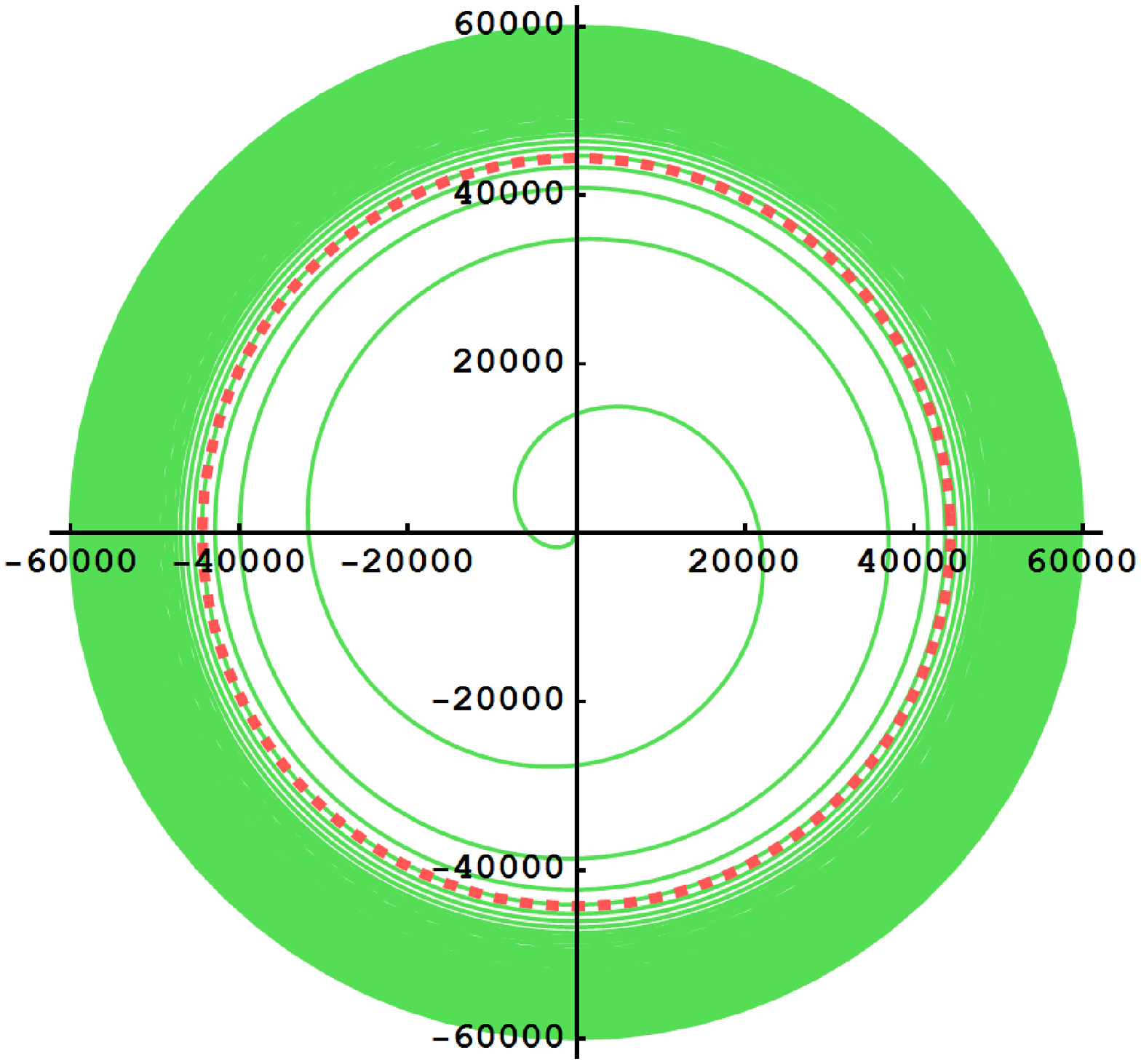}
\end{center}
\caption{
Numerical calculation of a quasi stable circular orbit of a massive particle
with $L / M = 62722$ in the $\tilde r$ - $\phi$ plane.
We set the initial radius of the circular orbit as
$\tilde r _s / M = 60000$ at $t_1 / t_0 = 10^9$.
In this set up, $T / t_0 = 1.1015 \times 10 ^6$,
then the condition $t_1 \gg T$ holds.
The radius $\tilde r_s$ and the period $T$ decreases gradually as the time $t_1$ increases.
At the time $t_1 / t_0 = (2 - \sqrt 3) L^2 / M^2 \simeq 1.0541 \times 10^9$,
the particle arrives at $\tilde r _{\rm qISCO} / M = 44351$ (red dotted circle),
then plunges into the black string. 
}
\label{fig:circle}
\end{figure}

Numerical calculations of the equations of motion~\eqref{E-L_eqs} shows the 
existence of quasi circular orbits (see Fig.~\ref{fig:circle}), and the particle in the orbit plunges 
into the black string. 

The period of quasi circular orbit is given by
\begin{align}
	T &= 2 \pi \frac{dt}{d\phi} 
= \sqrt{2} \pi  \sqrt{\frac{(\tilde M+\tilde r)^3 (\tilde M+2\tilde r)}{\tilde M \tilde r}} .
\end{align}
We can verify $t_1$ is much larger than $T$ in the numerical calculations. 
In the large $\tilde r$ limit, we have
\begin{align}
	T \to 2 \pi  \sqrt{\frac{\tilde r^3 }{ \tilde M }} .
\end{align}
It means the Kepler's third law.

\subsection{Massless particles with $p_w=0$}
Here, we consider a massless particle with $p_w=0$. 
We consider the four-dimensional metric 
\begin{align}
 ds^2 = a(\eta)^2\left[ -H^{-2} d\eta^2 + H \left( dr^2 + r^2 d\Omega^2_{\rm S^2} \right)  \right]  , 
\label{eq:metric4dim}
\end{align}
where we have used the conformal time $\eta=2\sqrt{t t_0}$.
Since a null geodesic is invariant under a conformal transformation, 
we can consider a null geodesic in the static metric 
\begin{align}
 d\bar{s}^2 =  -H^{-2} d\eta^2 + H \left( dr^2 + r^2 d\Omega^2_{\rm S^2} \right). 
\label{eq:metric4dimbar}
\end{align}
Since we can set $\theta=\pi/2$ without loss of generality,  
the Lagrangian for the massless particle in this metric is 
\begin{align}
 {\cal L} = \frac12\left( -H^{-2} \dot \eta^2 + H \left( \dot r^2 + r^2 \dot \phi^2 \right)\right), 
\label{masslessLag}
\end{align}
where the over dot denotes derivative with respect to an affine parameter. 
Conserved quantities are 
\begin{align}
 E = H^{-2} \dot \eta,\quad L=  H r^2 \dot \phi , 
\label{masslessL}
\end{align}
which are exactly conserved for massless particles. 
The massless condition is
\begin{align}
 H^{-1} \dot r^2   + U_{\rm eff} =  E^2, 
\label{massless_cond}
\end{align}
where 
\begin{align}
	U_{\rm eff}= \frac{L^2}{H^3 r^2 }.
\label{Ueff}
\end{align}
Therefore, the orbits are integrable in the case of $p_w=0$. 
By using the effective potential $U_{\rm eff}$, 
it is easy to know the existence of unstable circular null orbit as same as the 
Schwarzschild black hole. The radius is determined by
\begin{align}
	\frac{d U_{\rm eff}}{dr}= 0,  
\end{align}
as $r_n=M/2$. It is interesting that the unstable circular orbits are exact null geodesic  
solution even the original metric is time dependent. 
The \lq physical radius\rq\ $\tilde r_n=\tilde M/2=\sqrt{t_1/t_0} ~M/2$ increases in time, 
and $\tilde r_n < \tilde r_{\rm qISCO}$.  
The frequency of the photon received by a distant observer, whose proper time is $t$, 
is red shifted by the scale factor $a(t)$.

\section{Discussions towards stability analysis}\label{sec:solution2}

Since the present metric~\eqref{eq:metric3} describes a 
charged non-extremal black string,
one may expect that the spacetime has 
the Gregory-Laflamme instability~\cite{Gregory:1993vy}.
On the other hand, since the direction of an extra dimension
is shrinking as time increases in the solution,
one may also think that the spacetime 
will be stabilized at late time 
since the compactified extra dimension $w$ prohibits long wave length modes of 
perturbation.
We consider that the stability analysis for this spacetime is interesting,
but it is hard to study it since the background spacetime is time-dependent.
In this section, towards the stability analysis,
we give some discussions from two points of view.
One is that by focusing the late time geometry near the horizon which is 
approximately a static black string geometry,
we can study the stability problem on it.
The other is to consider the time evolution as a sequence of 
known static spacetimes with different physical parameters. 
Interestingly, each discussion suggests an opposite result.

\subsection{Gregory-Laflamme instability for the late time geometry near the horizon}

Near the horizon, the metric behaves approximately that of a static black string solution~\eqref{eq:metric4}
at late time.
In fact, this geometry can be obtained by taking
a special limit of the charged black string solution~\cite{Bleyer:1994wr} (see Appendix~\ref{appendix:hmblackstring}).
In~\cite{Frolov:2009jr}, Frolov and Shoom already showed the 
existence a Gregory-Laflamme instability for the charged black string solution for general parameters.
Surprisingly, the master equation for the zero mode gravitational perturbation 
only depends on a single parameter while the background spacetime has two parameters, {\it i.e.}, 
mass and charge parameters.
Thus, we can also expect that the late time geometry near the horizon, 
which can be obtained by a parameter limit,
also has a Gregory-Laflamme instability.

We should note that by taking the limit to the late time geometry near the horizon~\eqref{eq:metric4}
from the charged string solution with two parameters,
the asymptotic structure is also drastically changed as shown in Appendix.~\ref{appendix:hmblackstring}.
Before taking the limit, the spacetime is asymptotically flat 
towards the perpendicular direction to the black string,
but after the limit,
$t-r$ part of the metric asymptotes to AdS$_2$ and
the size of extra dimension at infinity becomes zero.
So, we should carefully check whether
the boundary conditions and the gauge conditions used in~\cite{Frolov:2009jr}
are still physically reasonable or not.
Fortunately, as shown in Appendix~\ref{appendix:stability},
such the difference in asymptotic structure 
does not affect on the derivation of the master equation and the boundary condition of it formally.
Thus, we can say that the late time geometry near the horizon~\eqref{eq:metric3} also has a Gregory-Laflamme mode
whose critical wave number along $w$ direction is given by
\begin{eqnarray}
 k_{\rm cr}  = 0.876\cdots,
\label{criticalwavenumber}
\end{eqnarray}
as like~\cite{Frolov:2009jr}.

\subsection{Time evolution as a sequence of static spacetimes}

Let us focus on the geometry near the 
time slice $t = t_1=
{\rm const.}$, 
the metric~\eqref{eq:metric3} behaves 
\begin{eqnarray}
ds^2 = - \left(1 + \frac{M}{r} \right)^{-2}
dt^2
+
\left(1 + \frac{M}{r} \right)
\left[
a(t_1)^2 (dr^2 + r^2 d\Omega_{\rm S^2}^2)
+
\frac{1}{a(t_1)^2}dw^2
\right].
\end{eqnarray}
Using the radial coordinate $\tilde{r}= a(t_1) r$, 
we find
\begin{eqnarray}
ds^2 = - \left(1 + \frac{\tilde M}{\tilde{r}} \right)^{-2}
dt^2
+
\left(1 + \frac{\tilde M}{\tilde{r}} \right)
\left[
d\tilde{r}^2 + \tilde{r}^2 d\Omega_{\rm S^2}^2
+
\frac{1}{a(t_1)^2}dw^2
\right].
\label{extremalhmbs}
\end{eqnarray}
This approximate geometry 
can be obtained by taking an extremal limit of
the charged black string solution~\cite{Bleyer:1994wr} (see Appendix~\ref{appendix:hmblackstring}).
In the extremal limit,
the mass parameter is $\tilde M$ and the size of extra dimension is $w_0 / a$
if $w$ direction is compactified with the period $w_0$.
We can consider that the 
time evolution is approximately described by increasing the scale factor $a$.
If the time scale of the instability is shorter than the Hubble scale,
we can expect that 
such a physical phenomenon is well described in this spacetime~(\ref{extremalhmbs}).
From this point of view, 
the mass parameter increases and the size of the extra dimension decreases as time increases.
This suggests that 
the spacetime will be stabilized at late time.\footnote{
In fact, the extremal limit of the charged black string
contains a curvature singularity on the horizon (see Appendix~\ref{appendix:hmblackstring}).
However, we can still use the result in~\cite{Frolov:2009jr}
as far as we consider a boundary condition where the perturbed quantiles do not 
diverge at the horizon and infinity.
We consider that such a boundary condition is physically reasonable now, 
since our original time-dependent solution~\eqref{eq:metric3} is regular on the horizon.}

\section{Summary and discussion} \label{summary}

We have investigated charged black string solutions residing in a five-dimensional Kasner universe where spatial three dimensions expand and an extra dimension contracts. The spacetime has an initial spatial singularity and admits an analytic extension across the event horizon.
The inner region is described by a time reversal of the outer region.
We can also generalize our solution to multi-black string and multi-black hole system.
In this system, the event horizon of a black string is still analytic
while both time-dependent and higher-dimensional multi black objects usually have non-smooth event 
horizon~\cite{Brill:1993tm, Gibbons:1994vm, Welch:1995dh, Candlish:2007fh, Kimura:2014uaa, Kanou:2014rya}.

Although there is no exact timelike Killing vector in the spacetime, the geometry is approximately static near the horizon.
In fact, the spacetime admits a unique second order asymptotic Killing generator
which satisfies an approximate Killing equation.
We can also see that the event horizon is an isolated horizon~\cite{Ashtekar:2004cn}.
The late time geometry near the horizon rapidly
approaches the geometry of a static exact charged black string solution.

We have studied motions of test particles in the spacetime
and seen that quasi stable circular orbits exist. If the time scale of cosmological evolution is much longer than that of orbiting particle near the black string,
we can effectively consider the scale factor to be constant during its typical motion.
From this point of view, we analyze the effective potential and show the existence of quasi circular orbits and quasi ISCO. The radius of a quasi circular orbit is slowly decreasing with time evolution,
then its orbit becomes an inspiral orbit.
This is because the energy of particle is not conserved due to the cosmological expansion.
After the particle reaches the quasi-ISCO radius, then the particle plunges into the horizon immediately.
If we focus on massless particles with $p_w = 0$, equation of motion reduces to that on a static black hole
since the spacetime is effectively conformal static.
We have shown that unstable circular orbits of photon exist as exact solutions.

We have given short discussions towards stability analysis from two approximate points of view.
One is that by focusing the late time geometry near the horizon which is approximately the geometry of a static charged black string solution,
we can discuss the stability problem.
This analysis suggests the existence of Gregory-Laflamme instability at late time.
On the other hand, since the direction of an extra dimension is shrinking,
we can also expect that the spacetime will be stabilized at late time
since the compactified extra dimension $w$ prohibits long wave length modes of perturbation.
By considering the time evolution as a sequence of static spacetime, we have obtained a suggestion that the spacetime will be stabilized at late time.
Since each discussion suggests an opposite result, the stability analysis of our spacetime without approximation is an interesting open question.
However, this is not an easy task because we need to study the perturbation around time dependent spacetime.
We leave this problem for future work.

Note added: D.~Klemm and M.~Nozawa study 
black holes in the expanding universe from the dimensional reduction
of supersymmetric solutions in (un)gauged supergravities~\cite{Klemm:2015qpi}.
They show that our solution can be transformed, via the
four-dimensional electromagnetic duality, into a supersymmetric
solution to the five-dimensional minimal ungauged supergravity.

\section*{Acknowledgments}

We thank D.~Klemm and M.~Nozawa for reading the
manuscript and sharing information before the submission.
We would like to thank G.~Gibbons, E.~Gourgoulhon, T.~Harada, T.~Houri, H.~Kodama, K.-i.~Nakao, and B.~Way for fruitful comments. 
This work is supported by the Grant-in-Aid for Scientific Research No.24540282. 
M.K. is supported by a grant for research abroad from JSPS.

\appendix

\section{Case of multi-black holes/strings} 
\label{extensionmulti}
The solution~\eqref{eq:metric3} can be easily generalized to multi-black string and multi-black hole solutions.   
The metric and the Maxwell field of such solutions are given by 
\begin{align}
ds^2 &= -H^{-2} dt^2 + H \left[ V ( dx^2 + dy^2 + dz^2 ) + V^{-1} ( dw + \bm \omega )^2 \right] , 
\label{eq:metric20}
\\
A_\mu dx^\mu &= \pm \frac{ \sqrt{3} }{ 2 } H^{-1} dt, 
\label{eq:field20}
\end{align}
where
\begin{align} 
 H &= 1 + \sum _i \frac{ M_i }{ |\bm x - \bm x _i| } ,
 \label{eq:h20}
\\
 V(t,r) &= \frac{ t }{ t_0} + \sum _i \frac{ N_i }{ |\bm x - \bm x _i| } ,
 \label{eq:v20}
\end{align}
the 1-form $\bm \omega $ is determined by $\nabla \times \bm \omega = \nabla V$, 
and $t_0,~ M_i ,~N_i$ are non-negative constants, and 
$\bm x = (x ,~ y ,~ z) ,~ \bm x _i = (x_i ,~ y_i ,~ z_i)$ denote a position vector
and a constant vector
on the three-dimensional flat Euclidean space.

Black string and black hole horizons are located at $t = \infty ,~ \bm x = \bm x _i$ 
with $t  |\bm x - \bm x _i| = {\rm finite}$.
The parameter $N_i$ controls the horizon topology of black strings and black holes.  
A point source $\bm x = \bm x _i$ with $M_i \neq 0 ,~ N_i = 0$ describes a dynamical black string 
discussed in the present paper, 
while a point source $\bm x = \bm x _i$ with $M_i \neq 0 ,~ N_i \neq 0$ describes a dynamical black hole 
discussed in~\cite{Kanou:2014rya}.  
Here,  
in the limit $t_0 \to \infty$ with $N_1 \neq 0$, otherwise $N_i = 0$,   
the metric~\eqref{eq:metric20} describes 
the five-dimensional extremal charged static asymptotically flat multi-black holes~\cite{Myers:1986rx}.  
On the other hand, 
introducing the new coordinate $t' = t - t_0$ then taking the limit $t_0 \to \infty$,  
the solution~\eqref{eq:metric20} reduces to 
five-dimensional static Kaluza-Klein multi-black hole solutions~\cite{Ishihara:2006iv}.

In the following, we show that the multi-black string solutions still admit analytic extensions
across the event horizons of the black strings.

\subsection{Extension across the event horizon in multi-black string system}
For simplicity, we restrict ourselves to the cases of two black strings, {\it i.e.}, 
$M_1 \neq 0 ,~ M_2 \neq 0$ and $N_1 = N_2 = 0$, otherwise $M_i = 0 ,~ N_i = 0$. 
Without loss of generality,  
we can put the locations of two point sources as 
$\bm x _1 = (0 , 0 , 0)$ and 
$\bm x _2 = (0 , 0 , \delta)$, 
where
the constant $\delta$ denotes the separation between two black strings. 
In this case, the metric and the Maxwell field are 
\begin{align}
ds^2 &= -H(r , \theta)^{-2} dt^2 + H(r , \theta) \left[ \frac{t}{t_0} \left( dr^2 + r^2 d\Omega^2_{\rm S^2} \right) 
+\frac{t_0}{t} dw^2 \right], 
\label{eq:metric}
\\
A_\mu dx^\mu &= \pm \frac{ \sqrt{3} }{ 2 } H(r , \theta)^{-1} dt, 
\label{eq:field}
\end{align}
where $H$ is given by  
\begin{align}
H(r , \theta) &= 1 + \frac{M_1}{r} + \frac{M_2}{\sqrt{r^2 + \delta ^2 - 2 \delta r \cos \theta}} . 
\label{eq:harmonich}
\end{align}

Similar to the single black string case, 
to extend the metric~\eqref{eq:metric} across the surface $r = 0 ,~ t = \infty$ with $r t = {\rm const.}$,  
we consider the null geodesics near the surface. 
The null condition is given by 
\begin{align}
-H(r , \theta)^{-2} dt^2 + H(r , \theta) \frac{t}{t_0} dr^2 = 0.
\end{align} 
Since, near $r = 0$, the function $H$ behaves as   
\begin{align}
H 
\simeq \frac{M_1}{r} + 1 + \frac{M_2}{\delta} ,
\end{align}
then we have 
\begin{align} 
\left( \frac{dt}{dr} \right)^2 = \frac{t}{t_0} \left( \frac{M_1}{r} + 1 + \frac{M_2}{\delta} \right) ^3 . 
\label{kinjinullcondition}
\end{align}

To obtain coordinates across the surface $r = 0 ,~ t = \infty$ with $r t = {\rm const.}$,  
we use an approximate ingoing future null geodesics near $r=0$  
\begin{align}
tr = \frac{1}{4 M_1 t_0} 
\left[ 
2 M_1 ^2 + u \sqrt{r M_1 \left( 1+\frac{M_2}{\delta} \right) } - 3 r M_1 \left( 1+\frac{M_2}{\delta} \right)  
\right] ^2 ,
\label{multiansatz}
\end{align} 
where $u$ denotes an arbitrary parameter which classifies the approximate null geodesics.

Using the curves~\eqref{multiansatz}, 
we introduce new coordinates $(u , \rho)$ as 
\begin{align}
r &= \frac{\delta \rho^2}{M_1 (M_2 + \delta)} ,
\\ 
t &= \frac{(M_2 + \delta) (2M_1 ^2 + \rho (u-3\rho))^2}{4 t_0 \delta \rho^2} ,
\end{align} 
then we rewrite the metric~\eqref{eq:metric} and the Maxwell field~\eqref{eq:field} 
in the $(u , \rho)$ coordinates as  
\begin{align} 
ds^2 = & \frac{(M_2 + \delta)^2 (2M_1 ^2 + \rho (u-3\rho))^2}{4 t_0 ^2 \delta^2 \rho^4 \tilde H^2}  
\left[
-\rho^2 du^2 
+ \rho^4 \left( \frac{4 \delta^3 \tilde H ^3}{M_1 ^2 (M_2 + \delta)^3} 
- \frac{(2M_1 ^2 + 3\rho^2)^2 }{\rho^6} \right) d\rho^2 
\right.
\notag \\
&
\left.
+ 2 (2M_1 ^2 + 3\rho^2) du d\rho 
+ \frac{\delta^3 \rho ^6 \tilde H ^3}{M_1 ^2 (M_2 + \delta)^3} d\Omega_{\rm S ^2} ^2 
+ \frac{16 t_0 ^4 \delta^3 \rho ^6 \tilde H ^3}{(M_2 + \delta)^3 (2M_1 ^2 + \rho (u-3\rho))^4} dw^2
\right] ,
\label{eq:metric2}
\\
A_\mu dx^\mu =& \pm \frac{\sqrt{3} (M_2 + \delta) (2M_1 ^2 + \rho (u-3\rho))}{4 t_0 \delta \rho^3 \tilde H} 
\left[ \rho^2 du - (2M_1 ^2 + 3\rho^2) d\rho \right] ,
\label{eq:field2}
\end{align} 
where 
\begin{align} 
\tilde H = 1+ \frac{M_1 ^2 (M_2 + \delta)}{\delta \rho^2} 
+ \frac{M_1 M_2 (M_2 + \delta)}{\delta \sqrt{\rho^4 + M_1 ^2 (M_2 + \delta)^2 - 2 M_1 (M_2 + \delta) \rho^2 \cos \theta} } .
\end{align}
We see that the metric~\eqref{eq:metric2} is analytic. 
When $M_2 = 0$, the metric~\eqref{eq:metric2} and the Maxwell field~\eqref{eq:field2} 
reduce to those in the single black string case.

In the limit $\rho \to 0$ with $u = {\rm finite}$  
(equivalently, $r \to 0 ,~ t \to \infty$ with $t r = M_1 ^3 / t_0$), 
we obtain 
\begin{align} 
ds^2 &\to \frac{4 M_1 ^2}{t_0 ^2} du d\rho + \frac{M_1 ^4}{t_0 ^2} d\Omega_{\rm S ^2} ^2 + \frac{t_0 ^2}{M_1 ^2} dw ^2.
\end{align} 
We find that the metric~\eqref{eq:metric2} 
is regular at the horizon $\rho = 0$.

Thus the solutions~\eqref{eq:metric} with~\eqref{eq:harmonich} describe 
a pair of charged black strings  
which have analytic event horizons without a singularity on the black string horizons 
in the five-dimensional Kaluza-Klein universe.

\section{Electrically charged black string with two parameters}
\label{appendix:hmblackstring}

We consider a static charged black string solution of five-dimensional Einstein-Maxwell equations
whose metric and gauge 1-form are given by
\begin{eqnarray}
ds^2 &=& - \left(1 - \frac{r_+}{r}\right)\left(1 - \frac{r_-}{r}\right)
dt^2
+
\left(1 - \frac{r_+}{r}\right)^{-1} dr^2
\notag \\ && 
\hspace{2cm}+ r^2 \left(1 - \frac{r_-}{r}\right) d\Omega^2_{\rm S^2}
+
\left(1 - \frac{r_-}{r}\right)^{-1} dz^2,
\label{chargedstringstatic}
\\
A_\mu dx^\mu &=& \pm \frac{\sqrt{3}}{2} \frac{\sqrt{r_+ r_-}}{r} dt,
\end{eqnarray}
where $r_\pm$ are parameters with the inequality $r_- \le r_+$,
and the coordinate range outside the horizon is $r_+ < r < \infty$.
The event horizon locates at $r = r_+$ and the curvature singularity locates at $r=r_-$.
This solution is a special limit of the diatonic charged black string solution constructed by
Bleyer and Ivashchuk~\cite{Bleyer:1994wr}.
Physical properties of the solution~\eqref{chargedstringstatic} 
were studied in~\cite{Horowitz:2002ym,Frolov:2009jr}.

First, we consider the extremal limit of this solution, {\it i.e.}, $r_- \to r_+$. 
By introducing coordinates 
$\tilde{r} := r - r_+$ and $w=a z$, 
we see that the metric~\eqref{chargedstringstatic} takes the form of~\eqref{extremalhmbs}.
Note that 
a curvature singularity locates at the horizon $r = r_+ (= r_-)$ in this extremal case.

\begin{figure}[h]
\begin{center}
 \includegraphics[width=0.45\linewidth,clip]{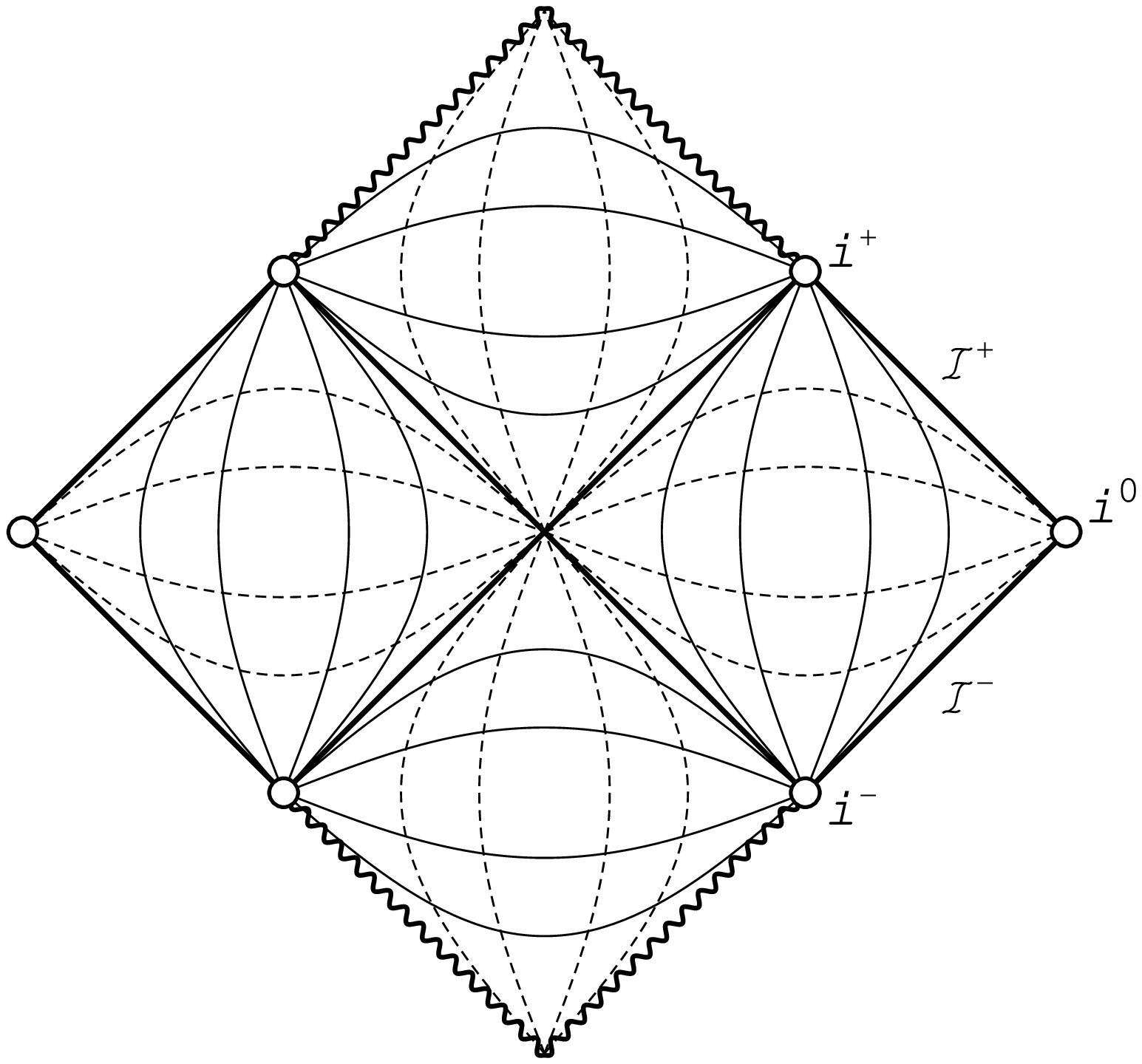}~~
 \includegraphics[width=0.25\linewidth,clip]{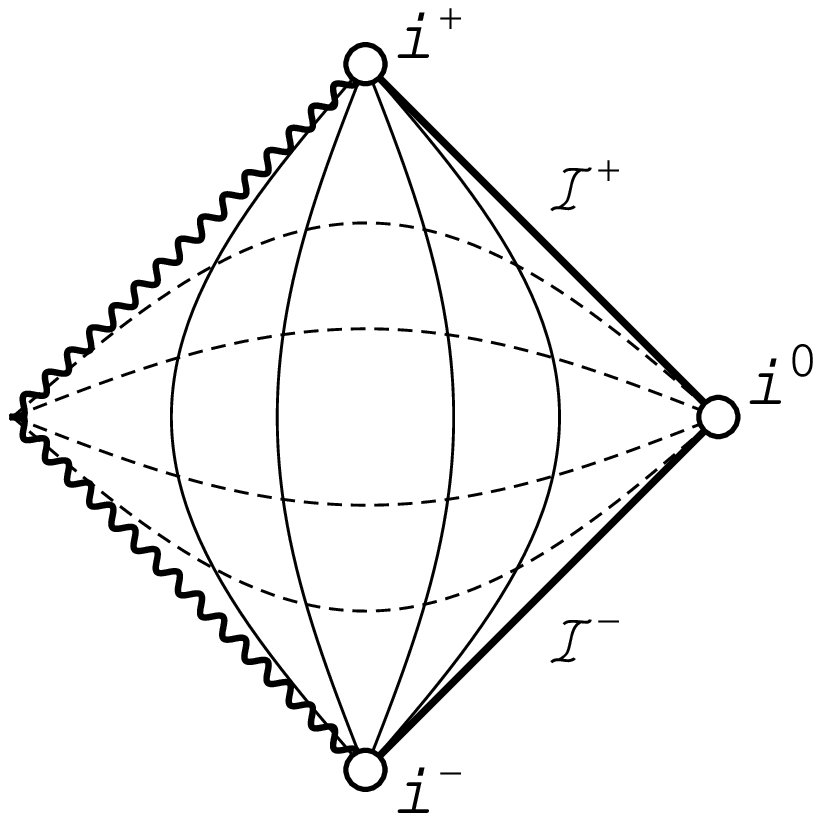}~~
 \includegraphics[width=0.25\linewidth,clip]{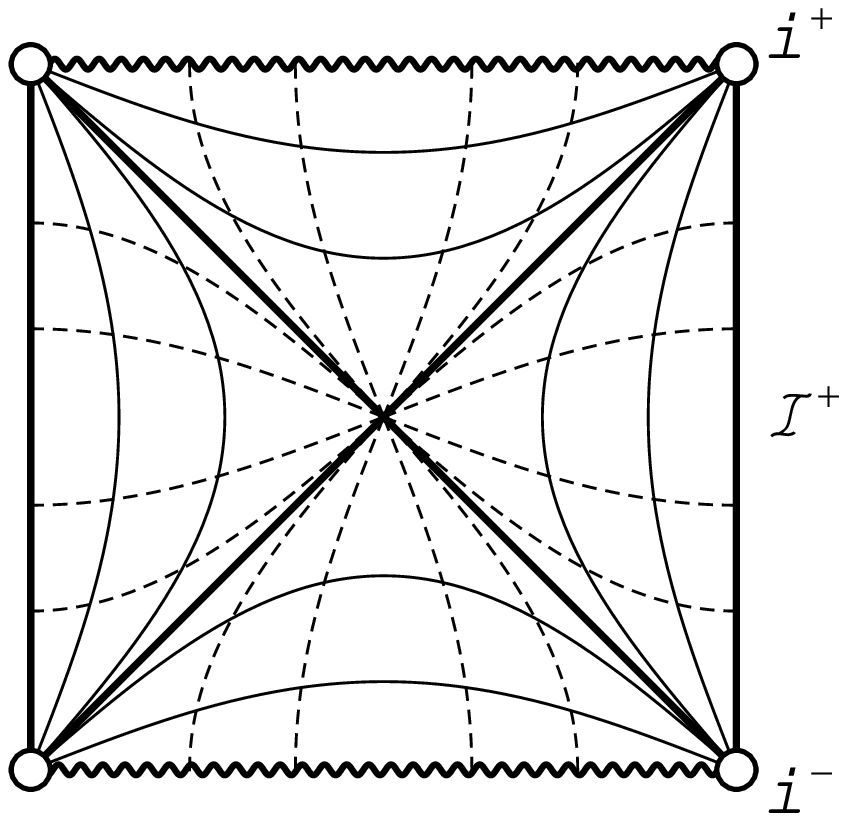}
\end{center}
\caption{Penrose diagrams of maximal extensions for static charged black strings.
Each figure corresponds to charged black strings with two parameters~(left), 
extremal limit~(center), $r_- \to \infty$ limit with the metric~\eqref{chargedstringstatic2} (right), respectively.
The wavy lines are curvature singularities. 
Dashed curves denote $t ={\rm const.}$~(left, center) or $T = {\rm const.}$~(right) surfaces, and 
thin solid curves denote $r ={\rm const.}$~(left, center) or $R = {\rm const.}$~(right) surfaces.}
\label{fig:penrose2}
\end{figure}

Second, we consider a special limit to obtain the late time geometry near the horizon.
Let us introduce new coordinates 
\begin{eqnarray}
T :=  \frac{R_h^2 }{r_- \sqrt{r_-^2 - R_h^2 }} t,~~
R^2  :=  r_-^2 \left(1 - \frac{r_-}{r}\right),~~
W :=  \frac{r_- z}{R_h},
\end{eqnarray}
with $R_h^2  :=  r_-^2 (1 - r_-/r_+)$,
then the metric~\eqref{chargedstringstatic} becomes
\begin{align}
ds^2 = - \frac{R^2 (R^2 - R_h^2)}{R_h^4} dT^2
+
\frac{4R^2}{R^2 - R_h^2} \frac{1 - R_h^2/r_-^2}{(1 - R^2/r_-^2)^4} dR^2
+
 \frac{R^2}{(1 - R^2/r_-^2)^2}d\Omega^2_{\rm S^2}
+
\frac{R_h^2}{R^2} dW^2, 
\label{chargedstringstatic2}
\end{align}
where 
the coordinate range outside the horizon is $R_h < R < r_- $. 
If we take a limit $r_- \to \infty$ while keeping $R_h = {\rm finite}$ for this metric,
we obtain the late time geometry near the horizon~\eqref{eq:metric4}.
In this limit, we can see that 
$T$-$R$ part of the metric asymptotes to AdS$_2$ near the infinity
and the size of the extra dimension at infinity is zero. 
We show the Penrose diagrams for these static charged black strings in Fig.~\ref{fig:penrose2}.

\section{Master equation for zero mode in the late time geometry near the horizon}
\label{appendix:stability}

In ref.~\cite{Frolov:2009jr}, it is shown that
gravitational and electric zero-mode perturbations are decoupled
for the static charged black string solution~\eqref{chargedstringstatic}, 
and only gravitational perturbation has unstable modes.
Since it is expected that the late time geometry near the horizon has a similar property,
we take the same ansatz for the metric and Maxwell field as~\cite{Frolov:2009jr},
\begin{eqnarray}
ds^2 &=& - \frac{R^2 (R^2 - R_h^2)}{R_h^4} e^{2 \tau(R,W) }dT^2
+
\frac{4R^2}{R^2 - R_h^2} e^{2 \sigma(R,W) }dR^2
+
R^2 e^{2 \gamma(R,W) } d\Omega_{\rm S ^2}^2
\notag\\&&
+
\frac{R_h^2}{R^2}  e^{2 \beta(R,W) }
(dW - \alpha(R,W) dR)^2, 
\\
A_{\mu}dx^\mu &=& \pm \frac{\sqrt{3}}{2} R^2 dT. 
\end{eqnarray}
By taking the first order of the small variables
$\tau(R,W), \beta(R,W), \sigma(R,W), \gamma(R,W), \alpha(R,W)$
for Einstein-Maxwell equations,
we can obtain the linearized equations.

Let us consider the Fourier transformation for $W$ direction, 
\begin{eqnarray}
&& 
\tau(R,W) = \tilde{\tau}(R) e^{ikW},~~
\sigma(R,W) = \tilde{\sigma}(R) e^{ikW},~~
\gamma(R,W) = \tilde{\gamma}(R) e^{ikW},~~
\notag
\\
&&
\alpha(R,W) = \tilde{\alpha}(R) e^{ikW}~~
\beta(R,W) = \tilde{\beta}(R) e^{ikW}.
\notag
\end{eqnarray}
By using the gauge transformations $g_{\mu \nu} \to g_{\mu \nu} - 2 \nabla_{(\mu}\xi_{\nu)}$,
the perturbed quantities behave as
\begin{eqnarray}
\tilde{\tau}(R) &\to & \tilde{\tau}(R) - \frac{2 R^2 - R_h^2}{4R^3}\tilde{\xi}_R(R), 
\\
\tilde{\sigma}(R) &\to & \tilde{\sigma}(R) 
 - 
\frac{R_h^2}{4R^3}\tilde{\xi}_R(R)
- 
\frac{R(R^2 - R_h^2)}{4R^3}\tilde{\xi}^\prime_R(R),
\\
\tilde{\gamma}(R) &\to & \tilde{\gamma}(R)
- 
 \frac{R^2 - R_h^2}{4R^3}\tilde{\xi}_R(R),
\\
\tilde{\alpha}(R) &\to & \tilde{\alpha}(R)
+
ikR^2 \tilde{\xi}_R(R) 
+
2 R \tilde{\xi}_W(R)
+
R^2 \tilde{\xi}^\prime_W(R),
\\
\tilde{\beta}(R) &\to & \tilde{\beta}(R)
+
\frac{R^2 - R_h^2}{4R^3} \tilde{\xi}_R(R)
-
ikR^2 \tilde{\xi}_W(R).
\end{eqnarray}
We can choose the gauge condition
\begin{eqnarray}
\tilde{\tau}(R) = 0,~~
\tilde{\beta}(R) = 0.
\label{gaugecondition}
\end{eqnarray}
Note that this completely fixes the gauge degrees of freedom and we can choose this gauge condition 
even near the horizon and infinity.\footnote{
When we consider the gauge transformation~\eqref{gaugecondition},
$\tilde{\xi}_R(R)$ may diverge at most of the order of $R$ near infinity.
However linear perturbation is still good approximation.
For the condition where the linear perturbation is a good approximation,
we should impose that the perturbed quantities 
$\tilde{\tau}(R), \tilde{\sigma}(R), \tilde{\gamma}(R), \tilde{\alpha}(R), \tilde{\beta}(R)$ 
take finite value in $R_h \le R \le \infty$, 
and the gauge transformation~\eqref{gaugecondition} does not violate this. }
After some calculations, we obtain a single master equation for $\tilde{\gamma}(R)$ as
\begin{eqnarray}
\frac{R^2 - R_h^2}{4R^4} \tilde{\gamma}^{\prime \prime}(R) 
-
\frac{4 R^4 - 3 R^2 R_h^2 - 3 R_h^4}{4R^5(4R^2-3 R_h^2)}\tilde{\gamma}^{\prime}(R) 
+
\frac{2 R_h^2}{R^4 (4R^2 - 3 R_h^2)} \tilde{\gamma}(R) 
&=& k^2 \tilde{\gamma}(R),
\label{master_eq}
\end{eqnarray}
The relations among $\tilde{\gamma}(R)$ and the other variables are given by
\begin{eqnarray}
\tilde{\alpha}(R) &=& \frac{-2i
\Big[4(-R R_h^2 + k^2 (4 R^7 - 3 R^5 Rh^2)) \tilde{\gamma}(R)
+
R_h^2 (2R^2 - 3 R_h^2) \tilde{\gamma}^\prime(R)\Big]
}{k(4R^2 - 3 R_h^2)^2},
\\
\tilde{\sigma}(R) &=& \frac{2(R^2 - R_h^2)(2 \tilde{\gamma}(R)+ R \tilde{\gamma}^\prime(R))}{4R^2 - 3 R_h^2}.
\end{eqnarray}
Introducing a new radial coordinate
\begin{eqnarray}
R^2 = \frac{R_h^2}{1-x},
\end{eqnarray}
where $x = 0$ and $x = 1$ correspond to the horizon and spatial infinity, respectively,
the master equation becomes 
\begin{eqnarray}
x^2 \frac{d^2\tilde{\gamma}(x)}{dx^2}
-
x
\frac{3x^2 + 6x-1}{(1-x)(1+3x)}
\frac{d\tilde{\gamma}(x)}{dx}
+ 
\left[
\frac{2x}{(1-x)(1+3x)}
-
\frac{x}{(1-x)^4}k^2 
\right]
\tilde{\gamma}(x)
= 0.
\end{eqnarray}
This completely coincides with the master equation in~\cite{Frolov:2009jr}.

The boundary conditions used in~\cite{Frolov:2009jr} are obtained by imposing that
the perturbed quantities do not diverge at the horizon and infinity.
Althogh the asymptotic structure of~\eqref{eq:metric4} 
 is different from that of the static charged black string~\eqref{chargedstringstatic},
the boundary conditions for the master equation~\eqref{master_eq}
that we should impose are the same, {\it i.e.}, 
the perturbed quantities do not diverge at $R=R_h$ and $R=\infty$.
Since we solve the same equation with the same boundary condition formally, 
we obtain the same result as~\cite{Frolov:2009jr}, {\it i.e.}, the spacetime is unstable against 
long wave perturbations whose critical wave number is $k_{\rm cr} =  0.876\cdots$.



\end{document}